# Acoustic Spectroscopy of Superfluid $^3$He in Aerogel

J.P. Davis, H. Choi, J. Pollanen, and W.P. Halperin

*Department of Physics and Astronomy, Northwestern University, Evanston, IL 60208, USA*

**Abstract:** We have designed an experiment to study the role of global anisotropic quasiparticle scattering on the dirty aerogel superfluid $^3$He system. We observe significant regions of two stable phases at temperatures below the superfluid transition at a pressure of 25 bar for a 98% aerogel.

## INTRODUCTION

Ultrasonic spectroscopy has proven to be a powerful tool in the study of $^3$He. The acoustic impedance for transverse sound exhibits anomalies at phase transitions that mark the superfluid phase diagram of $^3$He in 98% porosity silica aerogel [1]. The scattering of $^3$He quasiparticles from the silica aerogel strands suppresses $T_c$ and stabilizes the *B*-phase. An *A*-like phase is found to be metastable in zero field with large supercooling [1]. This is consistent with NMR [2] and low-frequency sound velocity measurements [3].

More recent acoustic tracking experiments by Vicente *et al.* [4] and NMR by Osheroff *et al.* [5] (99.3% aerogel) reveal that the *A*-like phase is in fact stable in a small temperature window near $T_c$ at high pressure. Vicente *et al.* suggest that this stabilization is due to the *local* anisotropic scattering from the aerogel strands. Furthermore, they propose introducing *global* anisotropy by uniaxial compression of the aerogel to study the effect.

## THEORY

Sauls [6] and Thuneberg *et al.* [7] have shown that local anisotropy can stabilize the axial state of superfluid $^3$He within aerogel. The relative stability of the axial (*A*) and isotropic (*B*) phases can be expressed as the difference between the beta parameters. The beta parameters are the coefficients of the fourth order terms in the Ginzburg-Landau expansion of the free energy in powers of the order parameter and are proportional to the difference in the heat capacity jumps. Sauls also noted [6] that large length scale correlations, or global anisotropy, in the aerogel might also favor phases with the orbital wavefunction perpendicular to the anisotropy axis, namely the planar or axial phases.

## EXPERIMENT

In order to study the role of anisotropy, one needs a probe that is both directional and extremely sensitive to phase transitions in the $^3$He. Transverse acoustic impedance has been shown to give a clear signature of all phase transitions in $^3$He [1]. The magnetic field dependence of the phase diagram allows us to assign which phases are equal spin pairing (ESP), like the *A*-phase or non-ESP, like the *B*-phase.

We designed and built a cell to compress a pair of aerogel samples that sandwich an *ac*-cut quartz acoustic transducer, as shown in Fig. 1. The electrical impedance was measured with a continuous wave impedance bridge [8]. A melting curve thermometer (MCT) was used as the primary thermometer.

The 98.2% aerogel in this experiment was grown at Northwestern using a two-step synthesis with rapid supercritical extraction (RSCE); and ~10% shrinkage was observed. Similar aerogels were studied by small-angle x-ray scattering (SAXS) as a function of compression [9]. Anisotropy increases systematically with uniaxial compression. Additionally, there is evidence of some intrinsic anisotropy [9].

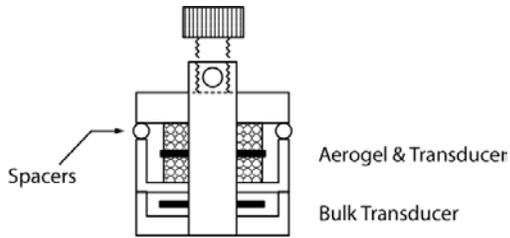

**FIGURE 1.** Compression cell. Spacers ensure transducer contact to the nominally uncompressed aerogel.

In our preliminary work, we performed temperature sweeps, Fig. 2, to determine how the $^3$He might be affected by this aerogel before removing the spacers and compressing the samples; this is the data that we present here.

## DATA AND RESULTS

At a pressure of 25 bar the bulk transition is at 2.36 mK, and is indicated by a separate bulk-transducer, Fig. 1. This trace is not shown in Fig. 2. We see no evidence for a bulk transition with the aerogel-sample transducer. In addition, the superfluid transition temperature in aerogel is less suppressed than previously observed by Gervais for a comparable porosity aerogel [1] ($T_{ca}$ = 1.91 mK). The transition from normal to superfluid appears to be in two parts, the superposition of a broad transition and a narrow transition. At lower temperatures there are also two distinct features in the acoustic impedance. On warming one of these is exceedingly sharp ($\Delta T \approx 2$ µK) and it exhibits a small hysteresis that can be associated with a first order transition. All of these features have been reproduced on multiple temperature sweeps.

The double transitions can most naturally be associated with there being two, non-identical, aerogel samples with which the transducer is in contact. Tentatively we associate the two low temperature features as transitions from *B* to *A*-like phases on warming, based on: a) previous studies of transverse impedance experiments [1], and b) their supercooling. The stability of the *A*-like phase might be a consequence of intrinsic global anisotropy [9], or possibly anisotropy introduced by nominal strain from the sample holder. Further work at different pressures and as a function of compression and magnetic field should help to clarify this situation.

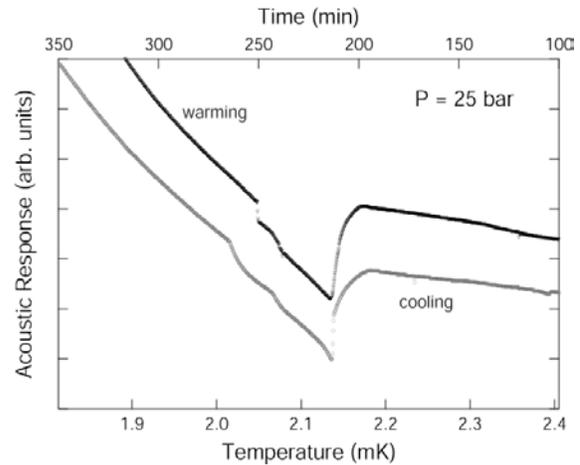

**FIGURE 2.** Acoustic impedance measurements of phase transitions for $^3$He within a 98.2% aerogel. The major feature is the transition to superfluid in aerogel; smaller features are discussed in the text.

## CONCLUSIONS

Preliminary studies of $^3$He at 25 bar in 98% aerogel grown at Northwestern suggest that an *A*-like phase can be stabilized, likely due to global anisotropy induced in the aerogel sample.

## ACKNOWLEDGMENTS

We would like to thank J.A. Sauls, N. Mulders, and Y. Lee for helpful discussions and acknowledge support from the National Science Foundation, DMR-0244099.